\documentclass[aps, pre, showpacs, twocolumn]{revtex4}

\usepackage{graphicx}
\usepackage{amsmath}
\graphicspath{{figs/}}

\newcommand{\be}{\begin{equation}}
\newcommand{\ee}{\end{equation}}

\newcommand{\rr}{\mathrm{R}}
\newcommand{\cm}{\mathrm{CM}}
\newcommand{\av}{\mathrm{av}}
\newcommand{\half}{{\textstyle \frac1{2}}}

\newcommand{\rma}{\mathrm{a}}
\newcommand{\rmb}{\mathrm{b}}

\begin{document}

\title{Model for Density Waves in Gravity-Driven Granular Flow in Narrow Pipes}

\date{\today}

\author{Simen {\AA}. \surname{Ellingsen}}
\affiliation{Department of Energy and Process Engineering,
Norwegian University of Science and Technology, N-7491 Trondheim,
Norway}
\author{Knut S. \surname{Gjerden}}
\affiliation{Department of Physics,
Norwegian University of Science and Technology, N-7491 Trondheim,
Norway}
\author{Morten \surname{Gr{\o}va}}
\affiliation{Department of Physics,
Norwegian University of Science and Technology, N-7491 Trondheim,
Norway}
\author{Alex \surname{Hansen}}
\affiliation{Department of Physics,
Norwegian University of Science and Technology, N-7491 Trondheim,
Norway}

\begin{abstract}
A gravity-driven flow of grains through a narrow pipe in vacuum is 
studied by means of a one-dimensional
model with two coefficients of restitution. Numerical simulations
show clearly how density waves form when a strikingly simple criterion is 
fulfilled: that dissipation due to collisions between the grains and 
the walls of the pipe is greater per collision than that which 
stems from collisions 
between particles. Counterintuitively, the highest flow rate is observed 
when the number of grains per density wave grows large. We find 
strong indication that the number of grains per 
density wave always approaches 
a constant as the particle number tends to 
infinity, and that collapse to a 
single wave, which was often observed also in previous simulations, occurs 
because the number of grains is insufficient for multiple wave formation.
\end{abstract}


\maketitle

\section{Introduction}

Transport of dry granular media through pipes and channels is a
problem of fundamental interest which is of considerable industrial
importance \cite{jackson00}, and it has been studied intensely both
experimentally and theoretically in recent decades
\cite{goodman71,jaeger96, jaeger92, goldhirsch03}.
Granular media behave radically different from both liquids and solids
\cite{jaeger96} and in granular pipe flow driven either by gravity or
pressurized gas, non-linear dynamical phenomena such as clogging
(density waves) are observed, but not yet well understood.

We consider dry grains falling inside a pipe. In this system,
encountered industrially, e.g., in emptying of silos and transportation
of sand or powder, there are three main mechanisms of interaction: (a)
collisions between grains, (b) collisions between grains and walls,
and (c) interaction of the grains with air in the system. Assuming the 
pipe is narrow, we approach
this problem by means of a simple one-dimensional model with periodic
boundary conditions in which collisions are modelled by means of two
coefficients of restitution $\mu$ and $\nu$, corresponding to the
collisions of mechanisms (a) and (b), respectively. We demonstrate that
this is sufficient to observe the formation of density waves.

% Nytt avsnitt 24.02.2010 - referanser til viktige simuleringer
Granular media has frequently been investigated by means of 
numerical simulations. Features such as clustering through dissipative 
collisions \cite{hopkins91, goldhirsch93} and inelastic collapse 
\cite{mcnamara92, mcnamara94} are among those reported.
Various driving mechanisms have been considered, which can roughly be 
divided into two categories: vibrating walls \cite{luding94, herbst04, herbst05} 
and interior heating \cite{cafiero00, williams96, peng98, vanzon04, vanzon05, bray07}. 
The case of an emptying hopper has been studied, see e.g.\ \cite{baxter89, risto94}, as has the decompaction transient 
associated with the release of an initially dense packing \cite{conrath00}.

\begin{figure*}[ht]
  \includegraphics[width=7in]{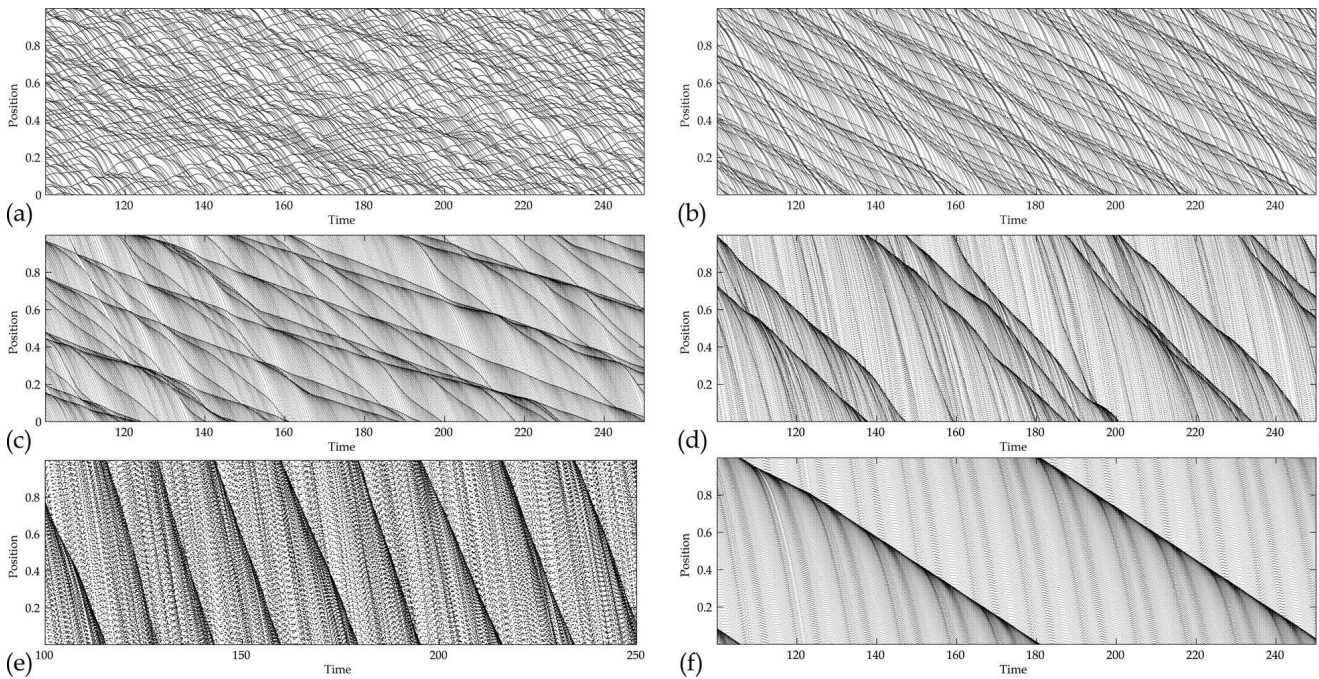}
  \caption{Spatiotemporal plots of grain trajectories. (a)
Gaseous regime $\mu=0.8, \nu=0.3$. (b) Split velocity
transition point $\mu=\nu=0.55$. (c) and (d): Different multi-wave
patterns, $(\mu,\nu)=(0.55,0.65)$ and $(0.30,0.85)$, respectively. (e)
and (f): Collapse to a single wave; $(\mu,\nu)=(0.70,0.95)$ and
$(0.15,0.85)$, respectively. Further details in text.}
  \label{bigfig}
\end{figure*}

% Experiment
The particular case of gravity-driven granular flow through channels 
and pipes has been studied for some time, both experimentally
and theoretically. In previous experimental investigations
\cite{moriyama98, reydellet00, bertho03, aider99, horikawa96, raafat96, bertho02} 
the importance of the presence of
air in the flow has been emphasized, and the formation of density
waves in falling sand has been explained with primary reference to the
air-grain interaction. Although clearly of importance in real systems, we
demonstrate that the introduction of air in the model is not necessary
to observe density waves. Our simulation suggests that
such flocculent behaviour, while certainly influenced by the presence
of air, can occur also in evacuated pipes, provided dissipation from
collisions between grains and walls is faster than
dissipation from grain-grain collisions.

% Simulation
Previous simulations of granular pipe flow have also reported density
waves \cite{hopkins91, poschel94, lee94, peng94, peng95, liss02, becco05}. 
Several mechanisms for velocity dissipation through
inelastic collisions, damping and static friction have been employed,
see, e.g., \cite{lee93}.
Early papers \cite{hopkins91, poschel94} were unable to study sufficiently 
large numbers of grains and collsions to reach a
state independent of initial conditions.
The model used by Lee \cite{lee94} was
a 2+1-dimensional time-driven molecular dynamics (MD) simulation, with
at least eight significant parameters.
In the model of Peng and Herrmann \cite{peng94,peng95}, a 2+1 dimensional lattice gas automaton is used in which
various events are assigned probalistic collision rules.
Another model was proposed by Liss, Conway and Glasser
\cite{liss02}. Again, this is a 2+1 dimensional model with more
complicated collision rules than ours. In their simulation the grain-wall
coefficient of restitution was found to be of little importance.
In all these models stable density waves were seen.

Our model is one-dimensional and much simpler, yet reproduces 
similar qualitative phenomena, dictated essentially by two
coefficients of restitution whose interpretation is physically transparent. 

%Outline
We present our model and the results of our
simulations and demonstrate that the model has two clearly distinct regimes
in the parameter plane, one in which density waves form (flocculent regime) and one
where no such are present (gaseous regime).
Average number of grains per density wave and flow
rate as functions of the two coefficients of restitution are
studied.

%%%%%%%%%%%%%%%%%%%%%%%%%%%%%%%%%%%%%%%%%%%%%%%%%%%%%%%
%%%%%%%%%%%%%%%%% S E C T I O N %%%%%%%%%%%%%%%%%%%%%%%
%%%%%%%%%%%%%%%%%%%%%%%%%%%%%%%%%%%%%%%%%%%%%%%%%%%%%%%
\section{The model}

The pipe of our model is sufficiently narrow to prevent grains from passing each other, and sufficiently wide to allow them to fall approximately freely between collisions. As a consequence, grain-wall collisions are triggered by, and occur immediately subsequent to, grain-grain collisions. The pipe is therefore assumed to be too narrow for arching effects to play a role. We thus propose that the system is essentially one-dimensional in behavior and may be captured qualitatively by a one-dimensional model where collisions with walls are incorporated into the grain-grain collision rules themselves.

The set-up is the following. Let $N$ grains move in one dimension
so that grain $i$ has position $x_i$ and velocity $v_i$. We use
periodic boundary conditions, with grains falling beneath $x=0$
reinserted at $x=L$, where $L$ is the length of the pipe. 
The grains are accelerated by a constant
gravity $g$ towards $x=0$. In the dilute limit $dN\ll L$, where $d$ is the diameter of a single grain, we find that the behaviour is independent of $d$, and we set $d=0$ for simplicity.

As mentioned, the collision rules employed are based on the idea that the falling
grains lose energy due to two different types of collisions, with the
walls and with each other. 
Grain-grain collisions are modelled by 
reducing the relative velocity of the colliding grains through
a coefficient of restitution 
$\mu\in [0,1]$. Thus, $v_{\rr,\rma\rmb}^+ =
-\mu v_{\rr,\rma\rmb}^-$ where $v_{\rr,\rma\rmb} = v_\rma-v_\rmb$, grain `$\rma$' is directly above grain
`$\rmb$', and we use superscripts `$-$' and `$+$' to denote times 
just before and after a collision, respectively.

Grain-wall collisions tend to decrease the average
velocity of the system, $v_\av=N^{-1}\sum_{i=1}^N v_i$, which would otherwise diverge in time. 
Instead, $v_\av$
eventually reaches and fluctuates around a
constant
value at which 
dissipated energy is in equilibrium
with energy gained from the gravitational field. We model the
interaction with the walls at each collision by a coefficient of
restitution $\nu\in [0,1]$ which dissipates grain energy by reducing the center of mass (CM)
velocity of the two colliding grains according to $v_{\cm,\rma\rmb}^+ = \nu
v_{\cm,\rma\rmb}^-$, where the CM velocity of the pair is 
$v_{\cm,\rma\rmb}=\half (v_\rma+v_\rmb)$. 
These rules combine to
\begin{subequations}\label{rules}
\begin{align}
  v_\rma^+ =& \half (\nu-\mu) v_\rma^- + \half (\nu+\mu) v_\rmb^- \\
  v_\rmb^+ =& \half (\nu+\mu) v_\rma^- + \half (\nu-\mu) v_\rmb^- .
\end{align}
\end{subequations}
Note that introducing a restitution of $v_\cm$, momentum is transfered for each collision 
from the grains to the (infinitely massive) pipe, and the momentum of the grains alone is 
not conserved. This causes the flow to approach a constant flow rate as observed in experiments.

While a collision changes the velocities of just two grains, the relative and CM 
velocities of three \emph{pairs} of grains are affected: the colliding pair plus 
the neighboring pairs above and below. In particular, a collision increases the 
relative velocity of the pair above the collision in proportion to the lost CM 
velocity of the colliding pair. Thus, relative velocities are constantly 
regenerated, and the phenomenon of inelastic collapse \cite{mcnamara92, mcnamara94} is 
avoided for $\mu>0$ and
$\nu<1$.

%%%%%%%%%%%%%%%%%%%%%%%%%%%%%%%%%%%%%%%%%%%%%%%%%%%%%%%
%%%%%%%%%%%%%%%%% S E C T I O N %%%%%%%%%%%%%%%%%%%%%%%
%%%%%%%%%%%%%%%%%%%%%%%%%%%%%%%%%%%%%%%%%%%%%%%%%%%%%%%
\section{Simulation and results.}

We use event-driven molecular dynamics for our simulations. 
Each grain is given an initial velocity
which is assigned randomly according to various initialization
schemes. The grains are also given equidistant starting positions between $0$ and $L$. 

Simulations are run for different grain numbers $N$ at constant grain
density $\rho=N/L$. After a `thermalization' time the system reaches a
steady state where the flow rate of the whole system fluctuates
about a mean value. 
Initializing the simulations with different initial conditions yields the
same dynamic steady state, characterized by average flow rate,
collision frequency and the average number of density waves.

In Fig.\ \ref{bigfig} we plot grain trajectories in space and time. In
these plots we have used, in arbitrary units, $L=1, g=
0.01$ and $N=100$. In these simulations initial velocities between $0$ 
and $1$ were drawn randomly from a uniform distribution.

We observe two distinct regimes in the $\mu,\nu$ plane. When $\nu<\mu$,
dense regions tend to spread out. In this `gaseous' regime, shown in
Fig.\ \ref{bigfig}a, no steady density waves are observed. Whenever $\nu>\mu$,
density waves are observed, and the transition from gaseous to flocculent behavior at
$\mu=\nu$ is sudden.

For the special case $\mu=\nu$ seen in Fig.\ \ref{bigfig}b, 
the eqs.\ (\ref{rules}) simplify to $v_\rma^+ = \mu v_\rmb^-$ and $v_\rmb^+ = \mu v_\rma^-$.
Now the velocity of grain `$\rma$'
after a collision is only a function of the velocity of grain `$\rmb$'
before the collision, and {\it vice versa}. After `thermalization' we
observe that the grain velocities are organized about two
values: a lower velocity for the grains whose last collision was with 
the grain below it and a higher velocity for those which last collided 
with the grain above it.

While the qualitative behaviour in the gaseous regime, $\mu>\nu$, is
insensitive to the values of $\mu$ and $\nu$,
the situation in the flocculent regime is quite different. As
exemplified for the parameter pairs in Fig.\ \ref{bigfig}, the wave
patterns vary strongly in this half of the parameter plane. Both
the magnitude and velocity of the density waves depend sensitively on
the values of $\mu$ and $\nu$. Moreover, the qualitative picture
varies from a few large and stable waves with approximately
constant velocity to many small and
volatile waves which emerge, merge and dissolve and vary greatly in
velocity even for a single set of parameters. Two examples
are shown in Fig.\ \ref{bigfig}c and d, but a multitude of different
multiwave patterns may be observed.

Beyond rescaling the time axis, 
the values of $L$ and $g$ do not affect the wave patterns,
so long as the grain number $N$ is large compared to the average
number of grains per density wave. For $N$ below some $\mu$,$\nu$-dependent threshold, the grains will gather in
one or a few stable or metastable waves, as exemplified in panels (e)
and (f) of Fig.\ \ref{bigfig}. 

\begin{figure}[tb]
  \includegraphics[width = 2.8in]{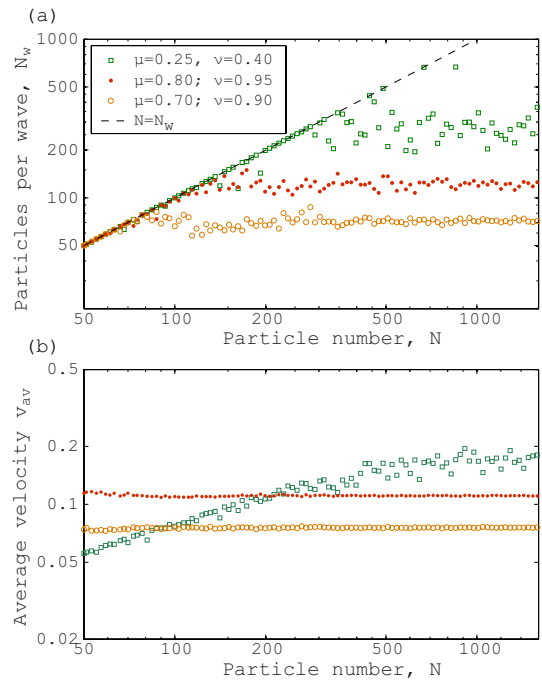}
  \caption{(Color online) (a) Number of grains per waves and (b) average grain velocity as functions of particle number $N$ for three sets of parameters (same in both panels). The dashed line in the above panel is $N=N_w$, i.e., a single density wave is present.}
  \label{fig:asymptotic}
\end{figure}

When $N$ is increased beyond this threshold, however, we find that the number of density waves
increases linearly, while the number of grains per density wave 
remain constant. Average flow rate and the average velocity of
density waves also remain independent of $N$ in the limit of
large $N$. This represents the asymptotic limit of our system, where there 
is no dependence on density $\rho$, which we illustrate for three pairs of coefficients in Fig.\ \ref{fig:asymptotic}. Due to limitations of computer time we have not 
investigated this limit when 
$N_w$ grows 
beyond $\approx 400$, which happens in the limit of either small $\mu$,
or $\nu\approx 1$, but we conjecture that such an 
asymptotic regime always exists when $\mu$ and $\nu$ are both on the open 
interval $\langle0,1\rangle$. Single-wave pictures such as panels (e) and (f) 
of Fig.~\ref{bigfig}, seen also in various other simulations \cite{liss02,peng94,peng95,lee94}, thus appear in our model simply because there are not enough particles in the system to form more than one wave \footnote{Naturally we cannot conclude that the same is necessarily true for the simulations of Refs.~\cite{liss02,peng94,peng95,lee94}.}.

\begin{figure}[tb]
  \includegraphics[width = 3.3in]{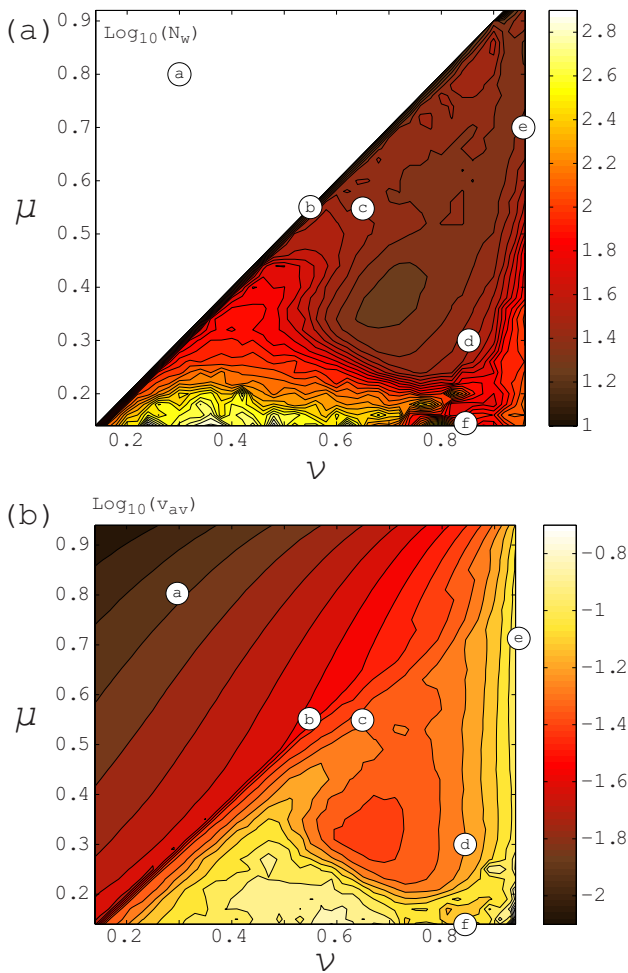}
  \caption{(Color online) (a): Plot of $\log N_w$ where $N_w$ is number of
grains per density wave as a function of $\mu$ and $\nu$ in the
asymptotic limit. (b): Plot of average flow rate as a function of
$\mu$ and $\nu$ in the asymptotic limit. In both panels, encircled letters refer to parameter pairs in Fig.~\ref{bigfig}.}
  \label{fig:FlowRate}
\end{figure}

In Fig.\ \ref{fig:FlowRate}a we estimate the
number of grains per wave in the flocculent regime, $N_w$, after
`thermalization' (calculated as
$N$ divided by the number of density waves) 
in the asymptotic (large $N$) regime. In
practice we choose $N$ large enough that the observed value of $N_w$
is constant with increasing $N$. We have used $N$ ranging from
$500$ to $4000$ in different areas of the plotted region. $N_w$
varies nonmonotonously and spans several orders of
magnitude. Indeed, when $\mu$ decreases below $0.2$ or $\nu$
approaches $1$, both $N_w$ and `thermalization' time diverge
rapidly, making calculation in this region expensive.

In Fig.\ \ref{fig:FlowRate}b the average flow rate is given for all values
of $\mu$ and $\nu$. In the gaseous regime we see a
monotonous variation supporting our observation that there is no
significant parameter dependency on the structures formed in this
region. Within the flocculent regime the
pattern is similar to that observed for the number of grains per density wave. 

Counterintuitively, flow rate is highest 
in areas of the $\mu,\nu$ plane in which the density waves are large, and conversely,
smaller waves correspond to a low flow rate. Since wave velocity is always lower than the flow rate, each grain will approach a density wave from above, collide its way through it, and fall into a low-density area beneath it once more before meeting another wave. 
%Although a single falling grain takes longer to penetrate a large density wave than a smaller one, 
Although large waves move more slowly than smaller waves,
we find that this is more than compensated by the presence of large low density areas between waves in which the grains can fall freely. Hence the total flow rate is governed by the length of these acceleration stretches whereas we find no clear connection between flow rate and wave velocity.

\begin{figure}[tb]
  \includegraphics[width = 3in]{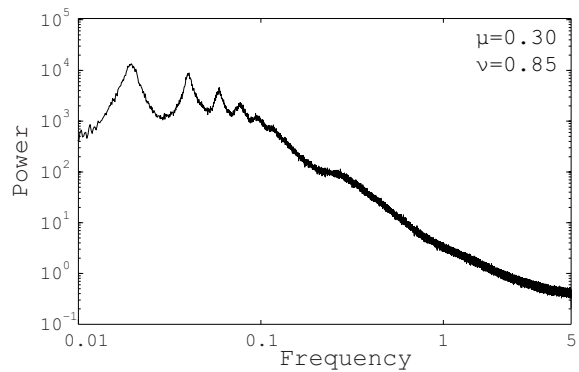}
  \caption{Frequency spectrum of the wave pattern of Fig.\ \ref{bigfig}d. 
  The graph is the average over the power spectra of 200 sub-samples of a 
  long time series of $\approx 10^7$ data points.}
  \label{fig:PowerSpectrum}
\end{figure}

We have analyzed the power spectrum of a typical wave pattern 
(that of Fig.\ \ref{bigfig}d) shown in Fig.\ 
\ref{fig:PowerSpectrum}. The analysis was performed by recording density in a 
region of length $L/10$ as a function of time. With sampling frequency $10$ per time unit, 
about $10^7$ data points were recorded. The series was then divided into 
subsamples which were Fourier transformed individually, and the average of 
the power for all sub-samples were then taken (different number of sub-samples were used for comparison; 200 sub-samples are used in figure \ref{fig:PowerSpectrum}). The spectrum is peaked at a 
few frequencies, demonstrating how the wave pattern has a number of preferred 
wave-front velocities. This is not obvious from studying spatiotemporal 
diagrams such as Fig.\ \ref{bigfig}d. In the high frequency regime the 
power dies off in a manner consistent with a power law with exponent -2,
consistent with the wave-fronts undergoing Brownian fluctuations.    

%%%%%%%%%%%%%%%%%%%%%%%%%%%%%%%%%%%%%%%%%%%%%%%%%%%%%%%
%%%%%%%%%%%%%%%%% S E C T I O N %%%%%%%%%%%%%%%%%%%%%%%
%%%%%%%%%%%%%%%%%%%%%%%%%%%%%%%%%%%%%%%%%%%%%%%%%%%%%%%
\section{Conclusions.}

We have studied granular pipe flow by means of a one-dimensional 
two-parameter model. The two parameters are the coefficients of 
restitution for collisions between grains and collisions between grains and 
walls. The very simple collision rules are contrasted by the 
amount of structure exhibited by the model, and formation of density waves 
varying greatly in magnitude and qualitative behavior is observed. We 
find a criterion for the formation of density waves: that the 
dissipation from collisions with walls be greater than that from 
grain-grain collision. Under this criterion our model predicts that 
density waves can form also in the absence of any interstitial gas.

Contrary to intuition, the flow rate is largest when density waves are 
large, slow and far between. 
This indicates that in some circumstances, the flow rate in gravity-driven granular pipe 
flow can be increased by softening or roughening the pipe walls.
For 
example, with soft grains described by $\mu=0.3$, a 'rough' pipe with $\nu=0.5$ gives a flow rate two to three 
times faster than a 'smoother' pipe for which $\nu=0.7$.

We have benefited from discussions with J.\ P.\ Hulin and the 
Granular Media Group at FAST laboratory, Orsay.

%%%%%%%%%%%%%%%%%%%%%%%%%%%%%%%%%%%%%%%%%%%%%%%%%%%%%%%
%%%%%%%%%%%%% B I B L I O G R A P H Y %%%%%%%%%%%%%%%%%
%%%%%%%%%%%%%%%%%%%%%%%%%%%%%%%%%%%%%%%%%%%%%%%%%%%%%%%

\end{document}